\documentclass[aps,prb,twocolumn,floatfix,groupedaddress,showpacs]{revtex4-1}

\usepackage{amssymb,amsmath}
\usepackage{color,soul}
\usepackage{graphicx}
\graphicspath{{./Figures/}}
\usepackage{hyperref}
\hypersetup{
     colorlinks   = true,
     citecolor    = blue,
     linkcolor   = blue
}

\newcommand{\0}{\phantom{0}}

\newcommand{\2}{\phantom{$(0)$}}
\newcommand{\+}{\phantom{+}}
\newcommand{\ms}{\mspace{-4mu}}
\newcommand{\TO}{\mathrm{TO}}
\newcommand{\LO}{\mathrm{LO}}

\begin{document}

\title{First-principles investigation of the structural, dynamical and dielectric properties of kesterite, stannite and PMCA phases of Cu$_2$ZnSnS$_4$}

\author{S. Poyyapakkam Ramkumar}
\email[]{sriram.ramkumar@uclouvain.be}
\affiliation{IMCN-NAPS, Universit\'{e} catholique de Louvain, Chemin des \'{E}toiles 8, B-1348 Louvain-la-Neuve, Belgium}
\affiliation{European Theoretical Spectroscopy Facility (ETSF)}
\author{Y. Gillet}
\affiliation{IMCN-NAPS, Universit\'{e} catholique de Louvain, Chemin des \'{E}toiles 8, B-1348 Louvain-la-Neuve, Belgium}
\affiliation{European Theoretical Spectroscopy Facility (ETSF)}
\author{A. Miglio}
\affiliation{IMCN-NAPS, Universit\'{e} catholique de Louvain, Chemin des \'{E}toiles 8, B-1348 Louvain-la-Neuve, Belgium}
\affiliation{European Theoretical Spectroscopy Facility (ETSF)}
\author{M.J. van Setten}
\affiliation{IMCN-NAPS, Universit\'{e} catholique de Louvain, Chemin des \'{E}toiles 8, B-1348 Louvain-la-Neuve, Belgium}
\affiliation{European Theoretical Spectroscopy Facility (ETSF)}
\author{X. Gonze}
\affiliation{IMCN-NAPS, Universit\'{e} catholique de Louvain, Chemin des \'{E}toiles 8, B-1348 Louvain-la-Neuve, Belgium}
\affiliation{European Theoretical Spectroscopy Facility (ETSF)}
\author{G.-M. Rignanese}
\email[]{gian-marco.rignanese@uclouvain.be}
\affiliation{IMCN-NAPS, Universit\'{e} catholique de Louvain, Chemin des \'{E}toiles 8, B-1348 Louvain-la-Neuve, Belgium}
\affiliation{European Theoretical Spectroscopy Facility (ETSF)}

\date{\today}

\begin{abstract}
Cu$_2$ZnSnS$_4$ (CZTS) is a promising material as an absorber in photovoltaic applications. The measured efficiency, however, is far from the theoretically predicted value for the known CZTS phases. To improve the understanding of this discrepancy we investigate the structural, dynamical, and dielectric of the three main phases of CZTS (kesterite, stannite, and PMCA) using density functional perturbation theory (DFPT). The effect of the exchange-correlation functional on the computed properties is analyzed. A qualitative agreement of the theoretical Raman spectrum with measurements is observed. However, none of the phases correspond to the experimental spectrum within the error bar that is usually to be expected for DFPT. This corroborates the need to consider cation disorder and other lattice defects extensively in this material.
\end{abstract}

\maketitle

\section{Introduction}

Photovoltaics are expected to play an important role in meeting the increasing demand of energy around the globe.
In order to do so in a sustainable way, solar cells composed of earth-abundant and non-toxic materials are required.
The absorber material, one of their key components, should have a band gap of $\sim$1.4-1.5~eV in order to achieve maximum efficiency (34\%) on the basis of Shockley-Queisser limit.~\cite{shockley1961detailed} 

In this context, Cu$_2$ZnSn(S,Se)$_4$ [CZT(S,Se)] has attracted considerable attention.
It has a high absorption coefficient $\sim$1$\times$10${^4}$ cm$^{-1}$.~\cite{seol2003electrical}
Unlike the current industrial state-of-the-art absorber Cu(In,Ga)(S,Se)$_2$ (CIGS), it does not contain any non-abundant element (such as indium or gallium).
Unfortunately, despite its band gap close to 1.5~eV,~\cite{matsushita2000thermal,scragg2008towards} the device efficiency is currently only 8.4\% for the pure sulfide CZTS~\cite{shin2013thin} and 12.6\% for a mixture of sulfur and selenium.~\cite{wang2014device} 

\begin{figure}[h]
\centering

\includegraphics{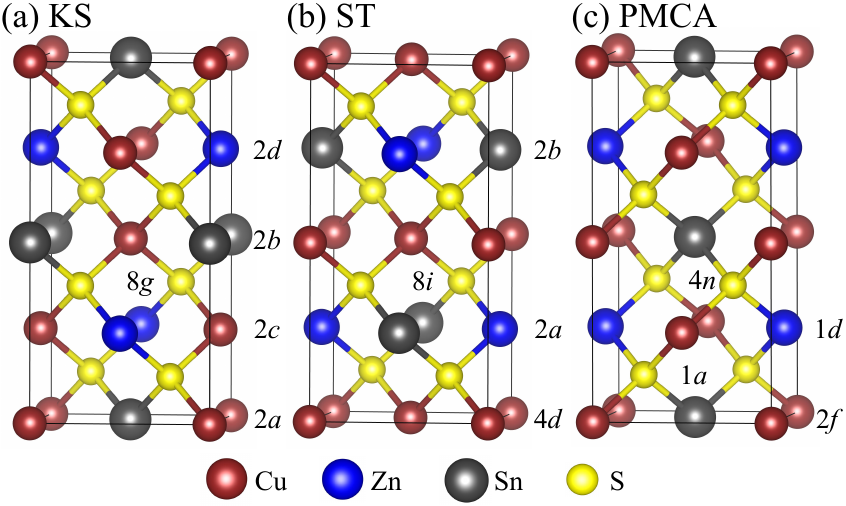}    
\caption{Conventional unit cells of the studied CZTS crystal structures: (a) Kesterite [KS], (b) stannite [ST], and (c) PMCA.
The Wyckoff positions of the Cu, Zn, Sn, and S atoms (in red, blue, grey, and yellow respectively) are also indicated.}

\label{fig:structs}
\end{figure}

The main factor limiting the efficiency is the open-circuit voltage, which is much lower than expected given the band gap of CZTS.~\cite{grossberg2014photoluminescence}
This could be due to the opening of multiple recombination paths, the presence of non-stoichiometric secondary phases (such as ZnS, Cu$_2$S, and Cu$_2$SnS$_3$), the coexistence of different stoichiometric structures or even disorder.~\cite{siebentritt2013kesterite,scragg2016cu}
Indeed, there are three main candidate structures for CZTS [kesterite (KS), stannite (ST) and primitive-mixed CuAu (PMCA), see Fig~\ref{fig:structs}].
First-principles calculations~\cite{chen2009crystal} show that KS is the more stable structure but the other two structures are only slightly higher in energy [$\Delta$E(ST)=2.9 meV/atom and $\Delta$E(PMCA)=3.2 meV/atom].
Furthermore, in KS, the energy needed to exchange Cu and Zn atoms in the planes at $z$=1/4 and 3/4 ($2c$ and $2d$ Wyckoff positions) has been computed to be fairly low,~\cite{chen2010defect} suggesting the existence of disorder in the Cu/Zn sublattice structure.
Experimental characterization of the synthesized structures using standard laboratory X-ray diffraction is difficult due to the similar atomic scattering factors of Cu and Zn.~\cite{charnock1996cu,cheng2011imaging,chen2010wurtzite}
Raman spectroscopy has also been considered for identifying the different structures or the possible disorder,~\cite{cheng2011imaging,khare2012calculation,dimitrievska2014multiwavelength,fontane2012vibrational,fernandes2011study} but the connection between a given structure and the corresponding Raman spectra is not straightforward at all.
Ideally, specific signatures in the spectra need to be identified for each possible structure in order to allow for a clear identification.

In this framework, first-principles calculations based on density functional theory (DFT)~\cite{hohenberg1964inhomogeneous} and density functional perturbation theory (DFPT)~\cite{giannozzi1991ab,gonze1997dynamical,gonze1997first} can be of considerable help.
In the case of CZTS, a few theoretical studies have already been performed.
The vibrational frequencies have been computed for the three different CZTS structures,~\cite{gurel2011characterization,khare2012calculation} but the complete Raman spectrum (with the intensities as well) has only been calculated for KS.~\cite{skelton2015vibrational}
Furthermore, the calculated results are known to depend on the functional adopted for modeling the exchange-correlation (XC) interactions.~\cite{he2014accuracy}
It is thus important to assess the significance of this dependence to enable the comparison with experiments.

In this study, the complete transmittance and Raman spectra of the three different CZTS crystal structures (KS, ST, and PMCA) are calculated and their dependence on the XC functional is evaluated.
We consider the local density approximation (LDA),~\cite{kohn1965self,oliveira2013born} the Perdew-Burke-Ernzerhof (PBE) functional,~\cite{perdew1996generalized} and its variant for solids (PBEsol).~\cite{perdew2008restoring}
The main conclusion emerging from this study is that the actual structure obtained in experiment is more complicated than the mere KS, ST, and PMCA structures or even a mixture of the three.  
For sake of completeness, the structural parameters, the Born effective charge tensors, and the dielectric constants computed using the three different functionals are also presented for the three CZTS structures.

\section{Computational details}

The structural, dynamical, and dielectric properties as well as Raman spectra are calculated with the \textsc{Abinit} software package.~\cite{gonze2002first,gonze2009abinit,gonze2016recent}
Norm-conserving pseudopotentials are used with Cu(3s$^2$3p$^6$3d$^{10}$4s$^1$), Zn(3s$^{2}$3p$^{6}$3d$^{10}$4s$^2$), Sn(4d$^{10}$5s$^2$5$p^2$) and S(3s$^2$3p$^4$) levels treated as valence states.
These pseudopotentials have been generated using the ONCVPSP package~\cite{hamann2013} and validated within the PseudoDojo framework.~\cite{lejaeghere2016reproducibility}
The wavefunctions are expanded using a plane-wave basis set up to a kinetic energy cutoff of 50~Ha for LDA (80~Ha for PBEsol and PBE).
The Brillouin zone is sampled using a 6$\times$6$\times$6 Monkhorst-Pack kpoint grid.~\cite{monkhorst1976special}
The different crystalline structures are fully relaxed with a tolerance of 1$\times$10$^{-8}$~Ha/Bohr ($\approx$5$\times$10$^{-4}$~meV/\AA) on the remaining forces.
The phonon frequencies are computed at the $\Gamma$ point of the Brillouin zone.
The Raman scattering efficiency of the phonon of frequency~$\omega_m$ for a photon of frequency~$\omega_i$ is defined as~\cite{veithen2005nonlinear}:
\begin{eqnarray}
I &=& (\omega_i - \omega_m)^4 | \textbf{e}_o . \boldsymbol{\alpha}_m . \textbf{e}_i |^2 \frac{n_m +1}{2 \omega^m}, \label{eq:Iquasi}
\end{eqnarray}
with $\textbf{e}_i$ and $\textbf{e}_o$ the incoming and outgoing light polarizations, 
$\boldsymbol{\alpha}^m$ the Raman susceptibility, $n_m$ the temperature-dependent phonon occupation factor:
\begin{equation}
n_m = \frac{1}{e^{{\omega_m}/{kT}}-1}.
\end{equation}
The Raman susceptibility  $\boldsymbol{\alpha}_m$ should be evaluated at the frequency of the incoming light, but is approximated in the present work by its static
value. Moreover, this susceptibility is obtained using DFPT~\cite{veithen2005nonlinear} (so without taking into account excitonic effects~\cite{Gillet2013})
and within the LDA (the corresponding PBEsol and PBE calculations are not yet available in \textsc{Abinit}).
The PBEsol and PBE Raman spectra thus are determined using the LDA intensities, assuming that the effect of the XC functional is weak for the intensities.
This approximation is justified on the following grounds.
First, the IR oscillator strengths calculated using LDA, PBEsol and PBE do not differ significantly.
Second, we have also computed the PBEsol Raman intensities using a finite-difference approach for a few selected phonon modes.
As reported in the Supplemental Material,~\cite{supplemental} we have found that they are very similar to those obtained within LDA both by the DFPT and finite-difference approaches.

We compute the transmittance from the absorbance and reflectance that are calculated from the real and the imaginary part of the dielectric functions ($\epsilon_1$ and $\epsilon_2$) for a thickness of 1 $\mu$m (a typical CZTS absorber material is between 1-2 $\mu$m) and phonon lifetime of 10 ps (a mid value estimated from the computational study of Skelton~\textit{et~al.}~\cite{skelton2015vibrational}).
The phonon lifetime ($\tau$) is related to the lorentzian broadening giving the full width at half maximum (FWHM or $\Gamma$) as $\tau$=1/2$\Gamma$.
The calculated transmittance along the parallel ($\parallel$) and perpendicular ($\perp$) directions to the $c$-axis of the conventional tetragonal cell are then averaged as $T_\mathrm{avg.}$=($2T_{\perp}$+$T_{\parallel}$)/3.

\section{Results and discussion}

\subsection{Structural properties}

The KS, ST and PMCA structures all contain 8 atoms in their primitive unit cell.
The conventional body-centered tetragonal unit cells of KS and ST with 16 atoms are shown in Fig.~\ref{fig:structs}.
To ease the comparison, the PMCA structure is also shown in the same figure, repeated twice along the $c$ axis.
All structures are composed of tetrahedra with two Cu, one Zn and one Sn atoms at the corners and a S atom close to the center so that the cations satisfy the octet rule of the S atom.
In fact, these CZTS phases can be obtained starting from a cubic zinc-blende (ZnS) cell through cross substitution of cations while preserving the octet rule of the S atom and the charge neutrality of the compound: four Zn atoms of the original ZnS structure are replaced by two Cu, one Zn and one Sn atoms.

In the KS structure (space group $I\bar{4}$, No. 82), there are two inequivalent Cu atoms occupying the Wyckoff sites $2a$ and $2c$; the Zn and Sn atoms are located at the $2b$ and $2d$ sites, respectively while the S atoms sit on $8g$ sites with three internal parameters ($u_x$,$u_y$,$u_z$).
In the ST structure (space group $I\bar{4}2m$, No. 121), all the Cu atoms are equivalent and reside in the $4d$ Wyckoff sites; the Zn and Sn atoms occupy the $2a$ and $2b$ sites, respectively; while the S atoms are located in the $8i$ sites with only two internal parameters ($u_x$,$u_x$,$u_z$).
In the PMCA structure (space group $P\bar{4}2m$, No. 111), all the Cu atoms sit on the $2f$ sites; the Zn and Sn atoms reside in the $1d$ and $1a$ sites, respectively, while the S atoms occupy $4n$ sites with two internal parameters ($u_x$,$u_x$,$u_z$).

The structural parameters (lattice constants and internal degrees of freedom) are optimized within DFT using the LDA, PBEsol, and PBE XC functionals.
The calculated values are reported in Table~\ref{tab:lattice-params}.
As typically found, LDA (resp. PBE) tends to underestimate (resp. overestimates) the lattice constants with respect to their experimental values.
PBEsol provides in-between values quite close to experiments.
Our results are consistent with previous calculations.~\cite{van1999correcting,skelton2015influence,he2014accuracy}
The effect of the XC functional on the lattice constants directly affects the different properties such as Born effective charge tensors, phonon frequencies, mode effective charge vectors and dielectric tensor computed in this paper.

\begin{table}
\caption{
\label{tab:lattice-params}
Lattice constants and internal parameters of CZTS in the KS, ST and PMCA phases computed with LDA, PBEsol and PBE.
The experimental data are taken from Ref.~\protect\onlinecite{hall1978kesterite}.
}
\centering
\begin{ruledtabular}
\begin{tabular}{l l c c c c c }
& & $a_0$ (\AA) & $c_0$ (\AA) & $u_x$ & $u_y$ & $u_z$ \\
\hline
     & Expt.  & 5.427 & 10.871 & 0.756 & 0.757 & 0.872\vspace{1mm}\\
KS   & LDA    & 5.316 & 10.636 & 0.761 & 0.770 & 0.870 \\
     & PBEsol & 5.367 & 10.785 & 0.760 & 0.769 & 0.870 \\
     & PBE    & 5.460 & 10.987 & 0.759 & 0.766 & 0.871\vspace{1mm}\\
ST   & LDA    & 5.316 & 10.637 & 0.760 &=$u_x$ & 0.886 \\
     & PBEsol & 5.367 & 10.784 & 0.760 &=$u_x$ & 0.885 \\
     & PBE    & 5.465 & 10.951 & 0.758 &=$u_x$ & 0.884\vspace{1mm} \\
PMCA & LDA    & 5.316 & 10.637 & 0.740 &=$u_x$ & 0.729 \\
     & PBEsol & 5.382 & 10.728 & 0.740 &=$u_x$ & 0.730 \\
     & PBE    & 5.473 & 10.930 & 0.742 &=$u_x$ & 0.732 \\
\end{tabular}
\end{ruledtabular}
\end{table}

In the following sections, we investigate the vibrational and dielectric properties of KS, ST and PMCA using DFPT within LDA, PBEsol and PBE.
For each functional, the corresponding optimized structure is used. 

\subsection{Born effective charge tensors}

We first compute the Born effective charge tensors of the Cu, Zn, Sn and S atoms for the three different structures.
For a given atom, its component $Z^*_{\alpha\beta}$ (with $\alpha,\beta$=$1,2,3$) is the proportionality coefficient relating, at linear order, the force on that atom in the direction $\alpha$ due and the homogeneous effective electric field along the direction $\beta$.
Equivalently, it also describes the linear relation between the induced polarization of the solid along the direction $\beta$ and the displacement of that atom in the direction $\alpha$, under the condition of zero electric field.~\cite{gonze1997dynamical}
The Born effective charge tensors obtained using LDA, PBEsol and PBE are shown in Table~\ref{tab:eff-charges} for all the atoms of KS, ST and PMCA.
For comparison, the nominal charges of the Cu, Zn, Sn and S atoms are 1, 2, 4 and -2, respectively.
Note that our LDA results for KS and ST are in good agreement with previous  calculations.~\cite{oliveira2013born} 

\begin{table*}[h]
\caption{
\label{tab:eff-charges}
Born effective charge tensors of the Cu, Zn, Sn and S atoms in KS, ST and PMCA calculated using LDA, PBESol and PBE. The eigenvalues of the symmetric part of the tensor are given in square brackets.
}
\setlength\tabcolsep{0.8pt}
\begin{ruledtabular}
\begin{tabular}{l@{\hspace{-4mm}}c@{\hspace{-1mm}}c@{\hspace{-1mm}}c@{\hspace{-1mm}}c}
& Z$^{*}_\mathrm{Cu}$ & Z$^{*}_\mathrm{Zn}$ & Z$^{*}_\mathrm{Sn}$ & Z$^{*}_\mathrm{S}$\vspace{2pt} \\
\hline
\multicolumn{4}{l}{LDA} \\
\ KS &
\begin{tabular}[t]{cc}
$\begin{pmatrix}
 +0.83 & \ms+0.42 &\ms\+0.00 \\
 -0.42 & \ms+0.83 &\ms\+0.00 \\
\+0.00 &\ms\+0.00 & \ms+0.60
\end{pmatrix}$
&
$\begin{pmatrix}
 +0.64 & \ms-0.09 &\ms\+0.00 \\
 +0.09 & \ms+0.64 &\ms\+0.00 \\ 
\+0.00 &\ms\+0.00 & \ms+0.82
\end{pmatrix}$\vspace{1mm}
\\
$\begin{bmatrix}
 +0.83 & \ms+0.83 & \ms+0.60
\end{bmatrix}$
&
$\begin{bmatrix}
 +0.64 & \ms+0.64 & \ms+0.82
\end{bmatrix}$\vspace{1mm}
\end{tabular}
&
\begin{tabular}[t]{c}
$\begin{pmatrix}
 +1.98 & \ms+0.35 &\ms\+0.00 \\
 -0.35 & \ms+1.98 &\ms\+0.00 \\
\+0.00 &\ms\+0.00 & \ms+2.13
\end{pmatrix}$\vspace{1mm} \\
$\begin{bmatrix}
 +1.98 & \ms+1.98 & \ms+2.13
\end{bmatrix}$\vspace{1mm}
\end{tabular}
&
\begin{tabular}[t]{c}
$\begin{pmatrix}
 +3.15 & \ms+0.08 &\ms\+0.00 \\
 -0.08 & \ms+3.15 &\ms\+0.00 \\
\+0.00 &\ms\+0.00 & \ms+3.24
\end{pmatrix}$\vspace{1mm}\\
$\begin{bmatrix}
 +3.15 & \ms+3.15 & \ms+3.24
\end{bmatrix}$\vspace{1mm}
\end{tabular}
&
\begin{tabular}[t]{c}
$\begin{pmatrix}
 -1.85 & \ms-0.43 & \ms+0.75 \\
 -0.05 & \ms-1.45 & \ms+0.09 \\
 +0.86 & \ms+0.15 & \ms-1.70
\end{pmatrix}$\vspace{1mm} \\
$\begin{bmatrix}
 -2.64 & \ms-1.41 & \ms-0.95
 \end{bmatrix}$\vspace{1mm}
\end{tabular}
\\
\ ST &
\begin{tabular}[t]{c}
$\begin{pmatrix}
 +0.96 & \ms+0.42 &\ms\+0.00 \\
 -0.42 & \ms+0.96 &\ms\+0.00 \\
\+0.00 &\ms\+0.00 & \ms+0.50
\end{pmatrix}$\vspace{1mm} \\
$\begin{bmatrix}
 +0.96 & \ms+0.96 & \ms+0.50
\end{bmatrix}$
\end{tabular}
&
\begin{tabular}[t]{c}
$\begin{pmatrix}
 +2.22 &\ms\+0.00 &\ms\+0.00 \\
\+0.00 & \ms+2.22 &\ms\+0.00 \\
\+0.00 &\ms\+0.00 & \ms+1.75
\end{pmatrix}$\vspace{1mm} \\
$\begin{bmatrix}
 +2.22 & \ms+2.22 & \ms+1.75
\end{bmatrix}$\vspace{1mm}
\end{tabular}
&
\begin{tabular}[t]{c}
$\begin{pmatrix}
 +3.13 &\ms\+0.00 &\ms\+0.00 \\
\+0.00 & \ms+3.13 &\ms\+0.00 \\
\+0.00 &\ms\+0.00 & \ms+3.19
\end{pmatrix}$\vspace{1mm} \\
$\begin{bmatrix}
 +3.13 & \ms+3.13 & \ms+3.19
\end{bmatrix}$\vspace{1mm}
\end{tabular}
&
\begin{tabular}[t]{c}
$\begin{pmatrix}
 -1.82 & \ms-0.86 & \ms+0.22 \\
 -0.86 & \ms-1.82 & \ms+0.22 \\
 -0.09 & \ms-0.09 & \ms-1.48
\end{pmatrix}$\vspace{1mm} \\
$\begin{bmatrix}
 -2.69 & \ms-1.47 & \ms-0.96
\end{bmatrix}$\vspace{1mm}
\end{tabular}
\\
\ PMCA &
\begin{tabular}[t]{c}
$\begin{pmatrix}
 +1.21 &\ms\+0.00 &\ms\+0.00 \\
\+0.00 & \ms+0.80 &\ms\+0.00 \\
\+0.00 &\ms\+0.00 & \ms+0.65
\end{pmatrix}$\vspace{1mm} \\
$\begin{bmatrix}
 +1.21 & \ms+0.80 & \ms+0.65
\end{bmatrix}$
\end{tabular}
&
\begin{tabular}[t]{c}
$\begin{pmatrix}
 +2.19 &\ms\+0.00 &\ms\+0.00 \\
\+0.00 & \ms+2.19 &\ms\+0.00 \\
\+0.00 &\ms\+0.00 & \ms+1.71
\end{pmatrix}$\vspace{1mm} \\
$\begin{bmatrix}
 +2.19 & \ms+2.19 & \ms+1.71
\end{bmatrix}$\vspace{1mm}
\end{tabular}
&
\begin{tabular}[t]{c}
$\begin{pmatrix}
 +3.28 &\ms\+0.00 &\ms\+0.00 \\
\+0.00 & \ms+3.28 &\ms\+0.00 \\
\+0.00 &\ms\+0.00 & \ms+3.12
\end{pmatrix}$\vspace{1mm}\\
$\begin{bmatrix}
 +3.28 & \ms+3.28 & \ms+3.12
 \end{bmatrix}$\vspace{1mm}
\end{tabular}
&
\begin{tabular}[t]{c}
$\begin{pmatrix}
 -1.87 & \ms-0.92 & \ms-0.50 \\
 -0.92 & \ms-1.87 & \ms-0.50 \\
 -0.13 & \ms-0.13 & \ms-1.53
\end{pmatrix}$\vspace{1mm} \\
$\begin{bmatrix}
 -2.80 & \ms-1.69 & \ms-0.78
\end{bmatrix}$\vspace{1mm}
\end{tabular}
\\
\hline
\multicolumn{4}{l}{PBEsol} \\
\ KS &
\begin{tabular}[t]{cc}
$ \begin{pmatrix}
 +0.88 & \ms+0.45 &\ms\+0.00 \\
 -0.45 & \ms+0.88 &\ms\+0.00 \\
\+0.00 &\ms\+0.00 & \ms+0.67
\end{pmatrix}$
&
$\begin{pmatrix}
 +0.90 & \ms-0.09 &\ms\+0.00 \\
 +0.09 & \ms+0.90 &\ms\+0.00 \\
\+0.00 &\ms\+0.00 & \ms+0.97
\end{pmatrix}$\vspace{1mm}
\\
$\begin{bmatrix}
 +0.88 & \ms+0.88 & \ms+0.67
\end{bmatrix}$
&
$\begin{bmatrix}
 +0.90 & \ms+0.90 & \ms+0.97
\end{bmatrix}$\vspace{1mm}
\end{tabular}
&
\begin{tabular}[t]{c}
$\begin{pmatrix}
 +2.01 & \ms+0.36 &\ms\+0.00 \\ 
 -0.36 & \ms+2.01 &\ms\+0.00 \\
\+0.00 &\ms\+0.00 & \ms+2.20
\end{pmatrix}$\vspace{1mm} \\
$\begin{bmatrix}
 +2.01 & \ms+2.01 & \ms+2.20
\end{bmatrix}$
\end{tabular}
&
\begin{tabular}[t]{c}
$\begin{pmatrix}
 +3.15 & \ms+0.09 &\ms\+0.00 \\
 -0.09 & \ms+3.15 &\ms\+0.00 \\
\+0.00 &\ms\+0.00 & \ms+3.25
\end{pmatrix}$\vspace{1mm}\\
$\begin{bmatrix}
 +3.15 & \ms+3.15 & \ms+3.25
\end{bmatrix}$\vspace{1mm}
\end{tabular}
&
\begin{tabular}[t]{c}
$\begin{pmatrix}
 -1.91 & \ms-0.44 & \ms+0.76 \\
 -0.04 & \ms-1.47 & \ms-0.09 \\
 +0.89 & \ms+0.16 & \ms-1.75
\end{pmatrix}$\vspace{1mm} \\
$\begin{bmatrix}
 -2.69 & \ms-1.47 & \ms-0.96
\end{bmatrix}$\vspace{1mm}
\end{tabular}
\\
\ ST &
\begin{tabular}[t]{c}
$\begin{pmatrix}
 +1.06 & \ms-0.46 &\ms\+0.00 \\
 +0.46 & \ms+1.06 &\ms\+0.00 \\
\+0.00 &\ms\+0.00 & \ms+0.57
\end{pmatrix}$\vspace{1mm} \\
$\begin{bmatrix}
 +1.06 & \ms+1.06 & \ms+0.57
\end{bmatrix}$\vspace{1mm}
\end{tabular}
&
\begin{tabular}[t]{c}
$\begin{pmatrix}
 +2.29 &\ms\+0.00 &\ms\+0.00 \\
\+0.00 & \ms+2.29 &\ms\+0.00 \\
\+0.00 &\ms\+0.00 & \ms+1.76
\end{pmatrix}$\vspace{1mm} \\
$\begin{bmatrix}
 +2.29 & \ms+2.29 & \ms+1.76
\end{bmatrix}$\vspace{1mm}
\end{tabular}
&
\begin{tabular}[t]{c}
$\begin{pmatrix}
 +3.12 &\ms\+0.00 &\ms\+0.00 \\
\+0.00 & \ms+3.12 &\ms\+0.00 \\
\+0.00 &\ms\+0.00 & \ms+3.19
\end{pmatrix}$\vspace{1mm} \\
$\begin{bmatrix}
 +3.12 & \ms+3.12 & \ms+3.19
\end{bmatrix}$\vspace{1mm}
\end{tabular}
&
\begin{tabular}[t]{c}
$\begin{pmatrix}
 -1.83 & \ms-0.88 & \ms+0.23 \\
 -0.88 & \ms-1.88 & \ms+0.23 \\
 -0.10 & \ms-0.10 & \ms-1.52
\end{pmatrix}$\vspace{1mm} \\
$\begin{bmatrix}
 -2.74 & \ms-1.51 & \ms-0.97
\end{bmatrix}$\vspace{1mm}
\end{tabular}
\\
\ PMCA &
\begin{tabular}[t]{c}
$\begin{pmatrix}
 +0.88 &\ms\+0.00 &\ms\+0.00 \\
\+0.00 & \ms+1.25 &\ms\+0.00 \\
\+0.00 &\ms\+0.00 & \ms+0.71
\end{pmatrix}$\vspace{1mm} \\
$\begin{bmatrix}
 +1.25 & \ms+0.88 & \ms+0.71
\end{bmatrix}$\vspace{1mm}
\end{tabular}
&
\begin{tabular}[t]{c}
$\begin{pmatrix}
 +2.26 &\ms\+0.00 &\ms\+0.00 \\
\+0.00 & \ms+2.26 &\ms\+0.00 \\
\+0.00 &\ms\+0.00 & \ms+1.69
\end{pmatrix}$\vspace{1mm} \\
$\begin{bmatrix}
 +2.26 & \ms+2.26 & \ms+1.69
\end{bmatrix}$\vspace{1mm}
\end{tabular}
&
\begin{tabular}[t]{c}
$\begin{pmatrix}
 +3.28 &\ms\+0.00 &\ms\+0.00 \\ 
\+0.00 & \ms+3.28 &\ms\+0.00 \\ 
\+0.00 &\ms\+0.00 & \ms+3.10
\end{pmatrix}$\vspace{1mm}\\
$\begin{bmatrix}
 +3.28 & \ms+3.28 & \ms+3.10
\end{bmatrix}$\vspace{1mm}
\end{tabular}
&
\begin{tabular}[t]{c}
$\begin{pmatrix}
 -1.91 & \ms-0.93 & \ms-0.50 \\
 -0.94 & \ms-1.92 & \ms-0.50 \\
 -0.11 & \ms-0.11 & \ms-1.55
\end{pmatrix}$\vspace{1mm} \\
$\begin{bmatrix}
 -2.98 & \ms-1.41 & \ms-0.97
\end{bmatrix}$\vspace{1mm}
\end{tabular}
\\
\hline
\multicolumn{4}{l}{PBE} \\
\ KS &
\begin{tabular}[t]{cc}
$ \begin{pmatrix}
 +1.02 & \ms+0.48 &\ms\+0.00 \\
 -0.48 & \ms+1.02 &\ms\+0.00 \\
\+0.00 &\ms\+0.00 & \ms+0.83
\end{pmatrix}$
&
$\begin{pmatrix}
 +0.72 & \ms-0.10 &\ms\+0.00 \\
 +0.10 & \ms+0.72 &\ms\+0.00 \\
\+0.00 &\ms\+0.00 & \ms+0.86
\end{pmatrix}$\vspace{1mm}
\\
$\begin{bmatrix}
 +1.02 & \ms+1.02 & \ms+0.83
\end{bmatrix}$
&
$\begin{bmatrix}
 +0.72 & \ms+0.72 & \ms+0.86
\end{bmatrix}$\vspace{1mm}
\end{tabular}
&
\begin{tabular}[t]{c}
$\begin{pmatrix}
 +2.08 & \ms+0.40 &\ms\+0.00 \\ 
 -0.40 & \ms+2.08 &\ms\+0.00 \\
\+0.00 &\ms\+0.00 & \ms+2.33
\end{pmatrix}$\vspace{1mm} \\
$\begin{bmatrix}
 +2.08 & \ms+2.08 & \ms+2.33
\end{bmatrix}$
\end{tabular}
&
\begin{tabular}[t]{c}
$\begin{pmatrix}
 +3.08 & \ms+0.07 &\ms\+0.00 \\
 -0.07 & \ms+3.08 &\ms\+0.00 \\
\+0.00 &\ms\+0.00 & \ms+3.24
\end{pmatrix}$\vspace{1mm}\\
$\begin{bmatrix}
 +3.09 & \ms+3.08 & \ms+3.24
\end{bmatrix}$\vspace{1mm}
\end{tabular}
&
\begin{tabular}[t]{c}
$\begin{pmatrix}
 -2.92 & \ms-0.44 & \ms+0.77 \\
 -0.00 & \ms-1.52 & \ms-0.11 \\
 +0.91 & \ms+0.14 & \ms-1.84
\end{pmatrix}$\vspace{1mm} \\
$\begin{bmatrix}
 -2.79 & \ms-1.54 & \ms-1.04
\end{bmatrix}$\vspace{1mm}
\end{tabular}
\\
\ ST &
\begin{tabular}[t]{c}
$\begin{pmatrix}
 +1.25 & \ms-0.49 &\ms\+0.00 \\
 +0.51 & \ms+1.25 &\ms\+0.00 \\
\+0.00 &\ms\+0.00 & \ms+0.73
\end{pmatrix}$\vspace{1mm} \\
$\begin{bmatrix}
 +1.25 & \ms+1.25 & \ms+0.73
\end{bmatrix}$\vspace{1mm}
\end{tabular}
&
\begin{tabular}[t]{c}
$\begin{pmatrix}
 +2.44 &\ms\+0.00 &\ms\+0.00 \\
\+0.00 & \ms+2.44 &\ms\+0.00 \\
\+0.00 &\ms\+0.00 & \ms+1.75
\end{pmatrix}$\vspace{1mm} \\
$\begin{bmatrix}
 +2.44 & \ms+2.44 & \ms+1.75
\end{bmatrix}$\vspace{1mm}
\end{tabular}
&
\begin{tabular}[t]{c}
$\begin{pmatrix}
 +3.09 &\ms\+0.00 &\ms\+0.00 \\
\+0.00 & \ms+3.09 &\ms\+0.00 \\
\+0.00 &\ms\+0.00 & \ms+3.09
\end{pmatrix}$\vspace{1mm} \\
$\begin{bmatrix}
 +3.09 & \ms+3.09 & \ms+3.09
\end{bmatrix}$\vspace{1mm}
\end{tabular}
&
\begin{tabular}[t]{c}
$\begin{pmatrix}
 -2.00 & \ms-0.91 & \ms+0.23 \\
 -0.91 & \ms-2.00 & \ms+0.23 \\
 -0.12 & \ms-0.12 & \ms-1.58
\end{pmatrix}$\vspace{1mm} \\
$\begin{bmatrix}
 -2.91 & \ms-1.57 & \ms-1.09
\end{bmatrix}$\vspace{1mm}
\end{tabular}
\\
\ PMCA &
\begin{tabular}[t]{c}
$\begin{pmatrix}
 +1.39 &\ms\+0.00 &\ms\+0.00 \\
\+0.00 & \ms+1.05 &\ms\+0.00 \\
\+0.00 &\ms\+0.00 & \ms+0.87
\end{pmatrix}$\vspace{1mm} \\
$\begin{bmatrix}
 +1.39 & \ms+1.05 & \ms+0.87
\end{bmatrix}$\vspace{1mm}
\end{tabular}
&
\begin{tabular}[t]{c}
$\begin{pmatrix}
 +2.38 &\ms\+0.00 &\ms\+0.00 \\
\+0.00 & \ms+2.38 &\ms\+0.00 \\
\+0.00 &\ms\+0.00 & \ms+1.66
\end{pmatrix}$\vspace{1mm} \\
$\begin{bmatrix}
 +2.38 & \ms+2.38 & \ms+1.66
\end{bmatrix}$\vspace{1mm}
\end{tabular}
&
\begin{tabular}[t]{c}
$\begin{pmatrix}
 +3.25 &\ms\+0.00 &\ms\+0.00 \\ 
\+0.00 & \ms+3.25 &\ms\+0.00 \\ 
\+0.00 &\ms\+0.00 & \ms+2.99
\end{pmatrix}$\vspace{1mm}\\
$\begin{bmatrix}
 +3.25 & \ms+3.25 & \ms+2.99
\end{bmatrix}$\vspace{1mm}
\end{tabular}
&
\begin{tabular}[t]{c}
$\begin{pmatrix}
 -2.02 & \ms-0.95 & \ms-0.49 \\
 -0.95 & \ms-2.02 & \ms-0.49 \\
 -0.08 & \ms-0.08 & \ms-1.60
\end{pmatrix}$\vspace{1mm} \\
$\begin{bmatrix}
 -3.08 & \ms-1.49 & \ms-1.07
\end{bmatrix}$\vspace{1mm}
\end{tabular}
\end{tabular}
\end{ruledtabular}
\end{table*}

The form of the Born effective charge tensor depends on the atomic site symmetry.
In particular, off-diagonal elements appear as the latter deviates from the tetrahedral symmetry.
This is the case for the S atoms in all the structures.
In contrast, the tensors of the Zn and Sn (resp. Cu, Zn, and Sn) atoms are diagonal in ST (resp. PMCA). None of the elements are diagonal in KS.
In all the structures, off-diagonal elements appear only in the direction perpendicular to the $c$ direction except for the S atoms for which all the components are non-zero.

For Cu in KS and ST, there is considerable deviation from the nominal charge, which implies more electron transfer from Cu to S.
Since Sn occupies the $2b$ Wyckoff site in both KS and ST, their values are quite similar in the two structures apart from a small deviation in KS originating from the lower symmetry of the anion sub-lattice to which the Sn atoms are bonded.

\subsection{Phonon frequencies at $\Gamma$}

We also compute the phonon frequencies at the $\Gamma$ point of the Brillouin zone.
Group theory analysis predicts the following irreducible representation for KS: 
\begin{center}
$\Gamma$=$\underbrace{1B \oplus 1E}_{\text{Acoustic}} \oplus \underbrace{3A \oplus \underbrace{ 6B\oplus6E}_{IR}}_{\text{Raman}}$,
\end{center}  
for ST:
\begin{center}
$\Gamma$=$ \underbrace{1B_{2} \oplus 1E}_{\text{Acoustic}} \oplus \underbrace{2A_{1} \oplus 2B_{1} \oplus \underbrace{ 4B_{2}\oplus6E}_{IR}}_{\text{Raman}}  \oplus \underbrace{A_{2}}_{\text{Silent}} $,
\end{center} 
and for PMCA:
\begin{center}
$\Gamma$=$ \underbrace{1B_{2} \oplus 1E}_{\text{Acoustic}} \oplus \underbrace{2A_{1} \oplus B_{1} \oplus \underbrace{4B_{2}\oplus6E}_{IR}}_{\text{Raman}} \oplus \underbrace{2A_{2}}_{\text{Silent}}$.
\end{center}

In Table~\ref{tab:phonon-freq}, we compare the results obtained within LDA, PBEsol and PBE. 
Our LDA phonon frequencies are in good agreement with those reported by G{\"u}rel \textit{et~al.}~\cite{gurel2011characterization} for KS and ST.
The average (resp. maximum) deviation is 1.5~cm$^{-1}$ (resp. 9.5~cm$^{-1}$) for KS and 1.4~cm$^{-1}$ (resp. 9.5~cm$^{-1}$) for ST. 

\begin{table*}[h]
\caption{
\label{tab:phonon-freq}
Calculated $\Gamma$-point phonon frequencies (in cm$^{-1}$) of CZTS KS, ST and PMCA structures using LDA, PBEsol and PBE. The experimental Raman (from Refs.~\protect\onlinecite{dimitrievska2014multiwavelength} and~\protect\onlinecite{fontane2012vibrational}) and IR (from Ref.~\protect\onlinecite{himmrich1991far}) frequencies are also shown for comparison.
}
\begin{ruledtabular}
\begin{tabular}{c c c l l c c c l l c c c l l c c c}
\multicolumn{3}{c}{Expt.}& & \multicolumn{4}{c}{KS}& & \multicolumn{4}{c}{ST} & & \multicolumn{4}{c}{PMCA} \\  
\cline{1-3} \cline{5-8} \cline{10-13} \cline{15-18}
\multicolumn{2}{c}{Raman\footnote{The first column is from Ref.~\protect\onlinecite{dimitrievska2014multiwavelength} and the second one from Ref.~\protect\onlinecite{fontane2012vibrational}}} & \multicolumn{1}{c}{IR} 
& & Mode & LDA & PBEsol & PBE 
& & Mode & LDA & PBEsol & PBE
& & Mode & LDA & PBEsol & PBE\\
    287 & 287 &     & & $A(1)$    & 301.8 & 289.3 & 270.4 & & $A_1(1)$    & 305.3 & 295.3 & 281.0 & & $A_1(1)$    & 321.3 & 308.3 & 290.6 \\
    302 &     &     & & $A(2)$    & 306.5 & 294.2 & 273.0 & & $A_1(2)$    & 323.2 & 311.6 & 293.5 & & $A_1(2)$    & 331.0 & 320.5 & 303.9 \\
    338 & 337 &     & & $A(3)$    & 326.7 & 315.3 & 298.8 & & $A_2$\2     & 306.3 & 291.9 & 269.3 & & $A_2(1)$\2     &\073.4 &\069.5 &\067.0 \\
   \082 &\083 &\086 & & $B(\TO1)$ &\094.3 &\089.0 &\085.5 & & $B_1(1)$    &\090.8 &\084.7 &\078.7 & & $A_2(2)$    & 303.8 & 288.7 & 263.1 \\
        &     &     & & $B(\LO1)$ &\095.4 &\090.9 &\086.0 & &             &       &       &       & &             &       &       &       \\
   \097 &\097 &     & & $B(\TO2)$ & 105.9 & 101.5 &\095.3 & & $B_1(2)$    & 321.9 & 307.9 & 284.2 & & $B_1$\2     & 327.4 & 313.1 & 290.9 \\
        &     &     & & $B(\LO2)$ & 106.1 & 101.7 &\095.3 & &             &       &       &       & &             &       &       &       \\
    164 & 166 & 168 & & $B(\TO3)$ & 178.5 & 170.5 & 159.4 & & $B_2(\TO1)$ &\096.9 &\093.4 &\091.2 & & $B_2(\TO1)$ &\080.8 &\079.5 &\080.7 \\
        &     &     & & $B(\LO3)$ & 178.6 & 170.7 & 159.6 & & $B_2(\LO1)$ &\097.0 &\093.5 &\091.2 & & $B_2(\LO1)$ &\080.9 &\079.7 &\080.9 \\
    255 & 252 &     & & $B(\TO4)$ & 268.4 & 253.3 & 225.8 & & $B_2(\TO2)$ & 171.0 & 161.7 & 146.2 & & $B_2(\TO2)$ & 170.8 & 161.5 & 146.6 \\
    263 &     &     & & $B(\LO4)$ & 284.6 & 269.7 & 243.3 & & $B_2(\LO2)$ & 172.3 & 161.8 & 146.9 & & $B_2(\LO2)$ & 170.9 & 161.7 & 147.5 \\
    331 &     & 316 & & $B(\TO5)$ & 329.2 & 315.9 & 295.7 & & $B_2(\TO3)$ & 305.5 & 293.7 & 275.0 & & $B_2(\TO3)$ & 303.0 & 289.3 & 270.6 \\
        &     &     & & $B(\LO5)$ & 333.0 & 319.9 & 300.0 & & $B_2(\LO3)$ & 320.9 & 308.6 & 289.6 & & $B_2(\LO3)$ & 309.0 & 294.8 & 275.8 \\
    353 & 353 &     & & $B(\TO6)$ & 349.5 & 339.0 & 321.2 & & $B_2(\TO4)$ & 354.1 & 343.5 & 326.1 & & $B_2(\TO4)$ & 329.5 & 320.2 & 303.8 \\
    374 &     &     & & $B(\LO6)$ & 359.0 & 349.6 & 333.7 & & $B_2(\LO4)$ & 357.9 & 348.3 & 331.6 & & $B_2(\LO4)$ & 339.4 & 330.4 & 313.9 \\
   \068 &\066 &\068 & & $E(\TO1)$ &\081.8 &\078.4 &\075.5 & & $E(\TO1)$   &\076.5 &\073.6 &\072.4 & & $E(\TO1)$   &\076.5 &\074.3 &\073.5 \\
        &     &     & & $E(\LO1)$ &\081.8 &\078.4 &\075.6 & & $E(\LO1)$   &\076.7 &\073.7 &\072.4 & & $E(\LO1)$   &\076.6 &\074.4 &\073.5 \\
   \097 &\097 &     & & $E(\TO2)$ & 101.8 &\097.3 &\095.0 & & $E(\TO2)$   & 108.5 & 103.2 &\097.6 & & $E(\TO2)$   &\092.4 &\090.3 &\087.8 \\
        &     &     & & $E(\LO2)$ & 101.8 &\097.3 &\095.1 & & $E(\LO2)$   & 108.6 & 103.3 &\097.9 & & $E(\LO2)$   &\092.4 &\090.3 &\087.8 \\
    140 & 143 & 143 & & $E(\TO3)$ & 166.1 & 158.1 & 144.3 & & $E(\TO3)$   & 170.1 & 163.0 & 152.0 & & $E(\TO3)$   & 177.7 & 169.8 & 157.4 \\
    151 &     &     & & $E(\LO3)$ & 166.2 & 158.1 & 144.4 & & $E(\LO3)$   & 170.1 & 163.0 & 152.0 & & $E(\LO3)$   & 177.9 & 170.0 & 157.7 \\
        &     & 255 & & $E(\TO4)$ & 277.3 & 262.4 & 236.9 & & $E(\TO4)$   & 267.4 & 251.7 & 224.3 & & $E(\TO4)$   & 266.4 & 250.8 & 222.4 \\
    271 & 272 &     & & $E(\LO4)$ & 288.3 & 273.2 & 246.9 & & $E(\LO4)$   & 280.6 & 264.3 & 236.5 & & $E(\LO4)$   & 278.6 & 262.9 & 234.6 \\
    316 &     & 293 & & $E(\TO5)$ & 312.0 & 297.0 & 273.6 & & $E(\TO5)$   & 306.9 & 292.0 & 269.7 & & $E(\TO5)$   & 307.7 & 293.8 & 272.4 \\
        &     &     & & $E(\LO5)$ & 314.4 & 299.8 & 278.1 & & $E(\LO5)$   & 313.0 & 298.4 & 276.5 & & $E(\LO5)$   & 310.7 & 296.9 & 276.0 \\
    347 & 347 & 351 & & $E(\TO6)$ & 331.9 & 322.9 & 306.8 & & $E(\TO6)$   & 333.5 & 324.2 & 307.5 & & $E(\TO6)$   & 328.7 & 319.2 & 301.6 \\
    366 &     &     & & $E(\LO6)$ & 342.0 & 333.5 & 317.9 & & $E(\LO6)$   & 342.5 & 333.3 & 317.3 & & $E(\LO6)$   & 341.7 & 333.0 & 316.7 \\
\end{tabular}
\end{ruledtabular}
\end{table*}

\begin{figure*}

\includegraphics{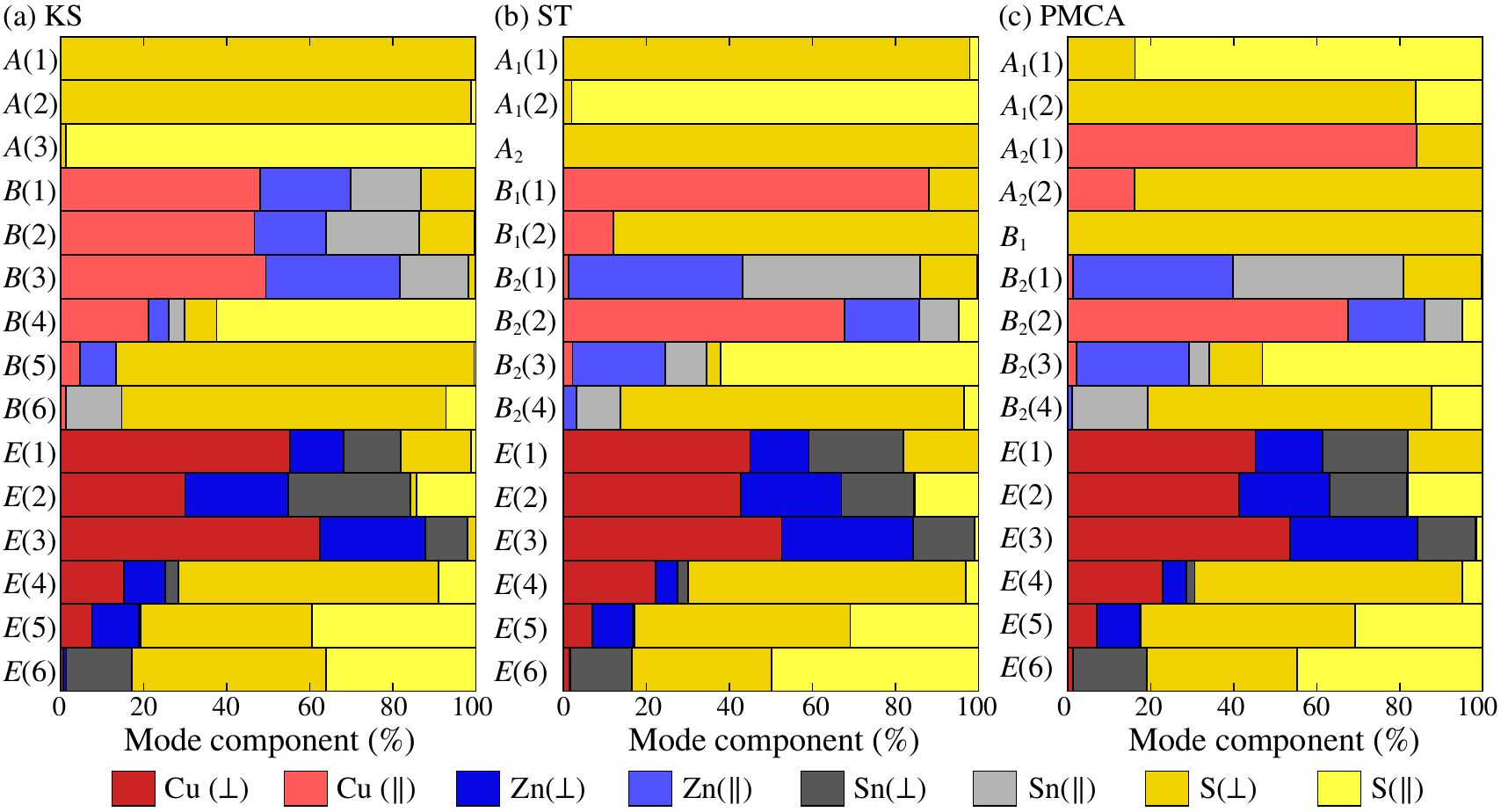}
\caption{
Atomic decomposition of the different vibrational modes as computed from LDA. The contribution from the Cu, Zn, Sn, and S atoms are shown in red, blue, grey, and yellow respectively.
The components parallel ($\parallel$) and perpendicular ($\perp$) to the $c$ axis are represented using light and dark shades of the colors associated to the different atoms.
}
\label{fig:decomp}
\end{figure*}

In order to analyze the atomic motion associated to the various modes, trying to highlight the similarities and differences among the different structures, we exploit the normalization condition of the eigendisplacements $U$ :
\begin{equation}
\sum_{\kappa\alpha} M_{\kappa}[U_{m}(\kappa\alpha)]^{*}U_{n}(\kappa\alpha)=\delta_{mn},
\end{equation}
where $M_{\kappa}$ is the mass of the ion $\kappa$, and $m$ and $n$ denote the phonon modes.
For a given mode, the contribution of each atom can be identified.
The results of such a decomposition obtained from calculations using LDA are presented in Fig.~\ref{fig:decomp} in which the components parallel ($\parallel$) and perpendicular ($\perp$) to the $c$ axis have also been separated.
In the figure, they are indicated using light and dark shades of the particular color associated to each atom. 
Note that, by using the eigendisplacements associated to two different structures in the formula, we can determine an overlap between the modes of the two phases. For sake of brevity, the three mode-by-mode overlap matrices (KS-ST, KS-PMCA, and ST-PMCA) are reported in Figs.S1-S3 of the Supplemental Material.~\cite{supplemental}


In KS, the $A$ modes only involve anionic motion (in yellow) occurring both in the direction $\parallel$ (light shade) and $\perp$ (dark shade) to $c$ axis.
The $B$ modes combine cationic motion parallel to the $c$ axis (in light red, blue, and gray for Cu, Zn, and Sn atoms, respectively) and anionic motion (in yellow).
Finally, the $E$ modes correspond to both cationic motion perpendicular to the $c$ axis (in dark red, blue, and gray) and anionic motion (in yellow).
Note that, for the $B$ and $E$ modes, the vibration of the S atoms happens both in parallel and perpendicular directions to the $c$ axis.

In ST, the $A_1$ and $A_2$ modes also only imply anionic motion (in yellow).
The $A_1(1)$ (resp. $A_2$) mode is very similar to the $A(1)$ [resp. $A(2)$] mode of KS.
This is confirmed by their strong overlap (see Fig.~S1 of the Supplemental Material.~\cite{supplemental})
These modes mainly involve anionic vibration perpendicular to the $c$ axis.
In contrast, the $A_1(2)$ mode essentially implies anionic vibration parallel to the $c$ axis (in light yellow).
It is very similar to the $A(3)$ mode of KS.
The $B_1$ modes correspond to the motion of Cu and S atoms parallel and perpendicular to the $c$ axis, respectively.
The $B_2$ modes combine the motion of Cu, Zn and Sn atoms parallel to the $c$-axis with the motion of S atoms.
Here, the strongest overlaps are (in decreasing order) between
the $B_2(4)$-ST and $B(6)$-KS modes (which mainly consist of the $\perp$ motion of S atoms and the $\parallel$ vibration of Sn atoms),
the $B_1(2)$-ST and $B(5)$-KS modes (which essentially involve the $\perp$ motion of S atoms and a minute contribution from the cations),
the $B_2(3)$-ST and $B(4)$-KS modes (corresponding to $\parallel$ motion of S atoms and a bit of cationic vibrations),
and the $B_2(2)$-ST and $B(1)$-KS modes (implying the $\parallel$ motion of the cations and a tiny anionic contribution).
The $B_1(1)$ and $B_2(1)$-ST modes correspond to a strong mixing of the $B(2)$ and $B(3)$-KS modes.
Finally, the $E$ modes consist of the motion of Cu, Zn and Sn atoms in the direction perpendicular to the $c$ axis and that of S atoms both in the $\perp$ and $\parallel$ directions.
The $E(4)$, $E(5)$ and $E(6)$-ST modes show strong one-to-one overlaps with the corresponding modes of KS, while $E(1)$, $E(2)$, and $E(3)$-ST modes are obtained by a strong mixing of the equivalent KS modes.

In PMCA, the $A_1$ modes involve only anionic motion.
The $A_1(1)$ [resp. $A_1(2)$] mode has a strong overlap with the $A_1(2)$ [resp. $A_1(1)$] mode of ST (see Fig.~S2 of Supplemental Material~\cite{supplemental}) and to a lesser extent the $A(3)$ [resp. $A(2)$] mode of KS (see Fig.~S3 of Supplemental Material~\cite{supplemental}).
The $A_2$ modes consist of the motion of Cu atoms parallel to the $c$~axis and of S atoms in the $\perp$ direction.
The $A_2(1)$-PMCA (resp. $A_2(2)$-PMCA) mode presents a strong overlap with the $B_1(1)$-ST (resp. $B_1(2)$-ST) mode.
For these modes, the correspondence with the KS modes is less obvious, the strongest overlap being between the $A_2(2)$-PMCA and $B(5)$-KS modes.
The $B_1$-PMCA mode involves only anionic motion in the direction perpendicular to the $c$ axis.
This mode is very similar to $A_2$-ST and is quite similar to the $A(1)$-KS mode.
The $B_2$ modes consist of both cationic motion parallel to the $c$ axis and vibration of the S atoms.
They present a strong one-to-one overlap with the $B_2$-ST modes.
Here also, the correspondence with the KS modes is more complicated apart from the strong overlap between the $B_2(4)$-PMCA and $B(6)$-KS modes. 
Finally, just like in KS and ST, the $E$ modes comprise both cationic motion in the direction perpendicular to the $c$ axis and motion of the S atoms in all directions.
For the first three $E$ modes, the one-to-one PMCA-ST overlap is much stronger than the KS-ST one.
In particular, the $E(1)$-PMCA and $E(1)$-ST modes are very much alike.
For the last three $E$ modes, the one-to-one PMCA-ST overlap is also quite significant, similarly to the KS-ST one.
The only difference is that the $E(5)$ and $E(6)$ modes of PMCA involve a stronger mixing of the corresponding ST modes.
Once again, the comparison between PMCA and KS modes is more involved. 
For all $E$ modes, the PMCA modes imply a strong mixing of KS modes, the strongest overlap being between the $E(4)$-PMCA and $E(4)$-KS modes.

\subsection{Transmittance and Raman spectra}

\begin{figure*}
\includegraphics{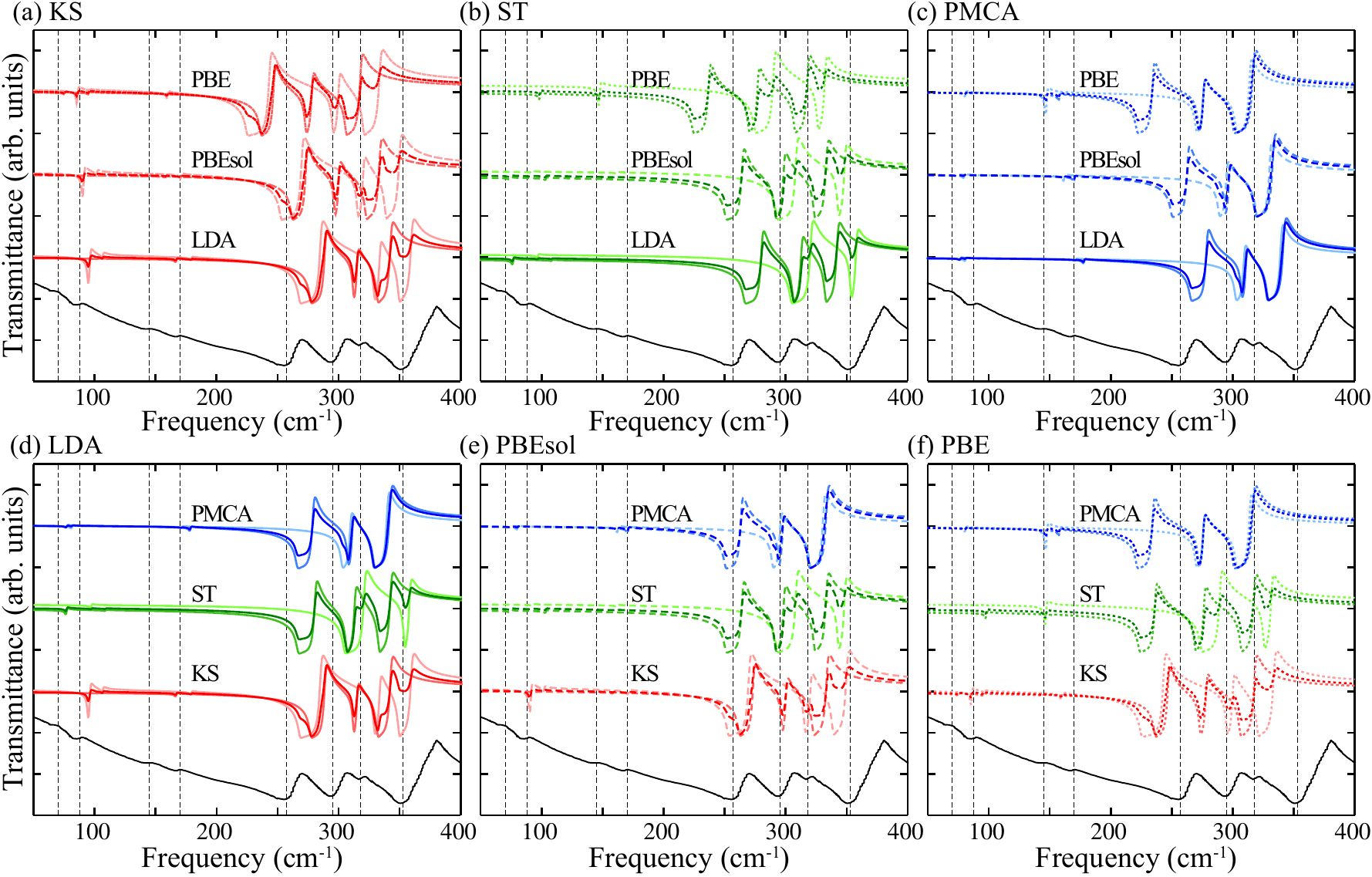}
\caption{
Transmittance calculated using LDA, PBEsol and PBE for KS, ST and PMCA.
For comparison with experiment~\cite{himmrich1991far}, the absorber thickness is taken to be 1 $\mu$m and the phonon lifetime to be 10 ps.
Two series of figures are presented:
panels (a), (b) and (c) allow for a comparison between the different functionals for KS, ST and PMCA (red, green, and blue curves, respectively);
panels (d), (e) and (f) permit to contrast the various phase with LDA, PBEsol and PBE (solid, dashed, and dotted curves, respectively).
The light, medium, and dark curves represent the parallel, perpendicular, and average transmittance, respectively (the directions being indicated with respect to the $c$ axis of the conventional tetragonal cell).
The experimental results are shown as a black curve at the bottom of each panel.
The black vertical dashed lines are drawn as a guide for the eyes at the frequencies measured experimentally.   }
\label{fig:ir}
\end{figure*}

The experimental far-IR absorption spectra (see Fig.~\ref{fig:ir}), which determines the transmittance, shows 4 peaks in the low frequency region ($<$200 cm$^{-1}$) and 4 peaks in the high frequency region ($>$200 cm$^{-1}$).
When considering the different theoretical results, we first note that the effect of the XC functional is mostly described as a shift along the frequency axis (a more quantitative discussion is given below for the Raman spectra).
This hence changes the quantitative agreement with the experimental spectrum but not the qualitative features.
In fact, all the computed (orientationally averaged) transmittance curves show a qualitative agreement with the experimental spectrum in the higher frequency region. Nevertheless, the calculated frequencies are not on top of the experimentally observed peaks and there are also some peaks seen in experiment but missing in the calculated spectra. In the lower frequency region, the same situation repeats with no unique matching of any one structure with the experimental spectrum. It is clear that none of the three different structures conforms with the experiment individually, nor does a combination of them. This indicates the need to consider other disordered structures to characterize the sample.

\begin{figure*}
\includegraphics{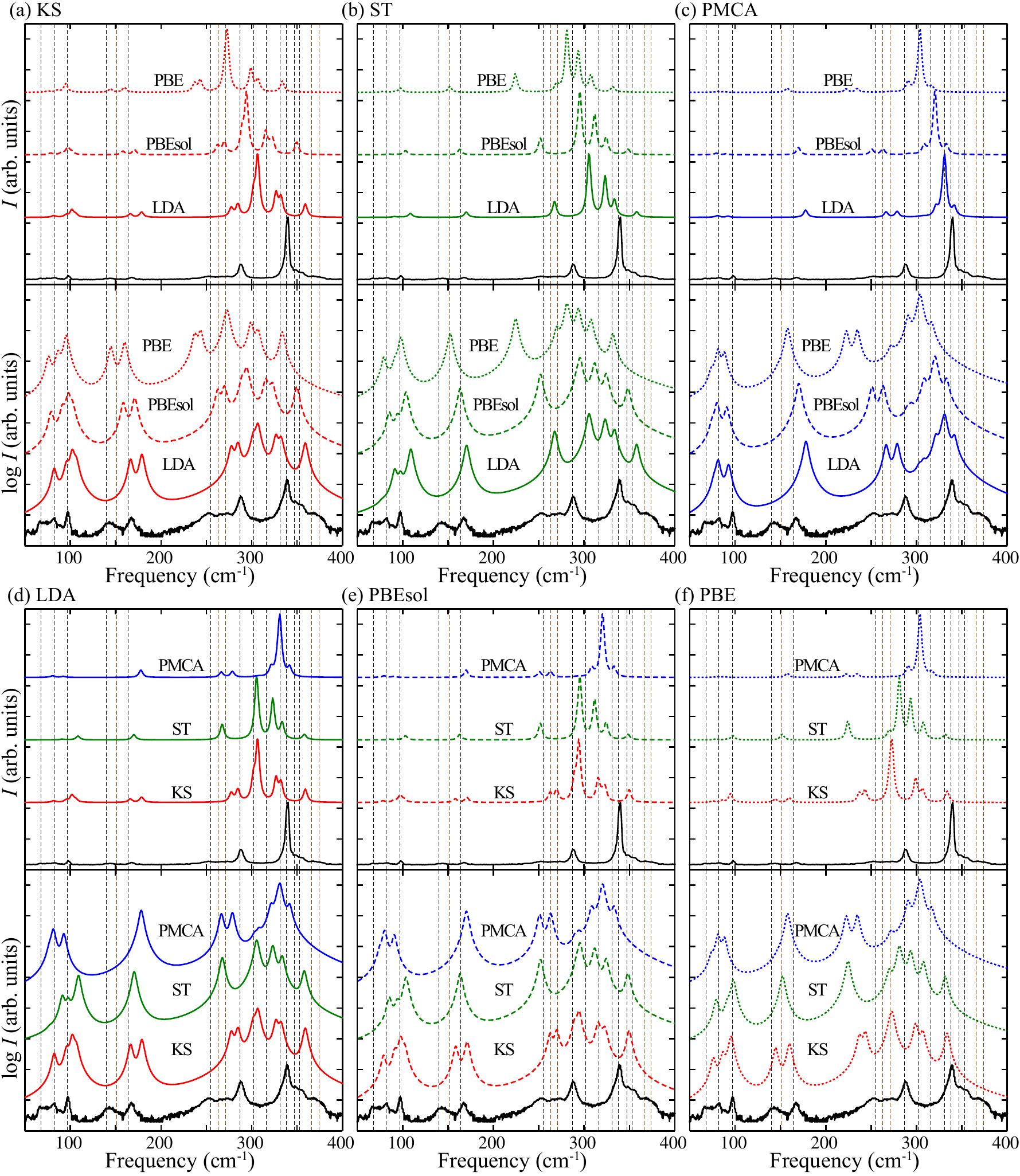}
\caption{
Raman spectra computed using LDA, PBEsol and PBE for KS, ST and PMCA. 
For comparison with experiment~\cite{dimitrievska2014multiwavelength}, a temperature of 300~K is chosen for the Bose-Einstein occupation factors.
Two series of figures are presented:
panels (a), (b) and (c) allow for a comparison between the different functionals for KS, ST and PMCA (red, green, and blue curves, respectively);
panels (d), (e) and (f) permit to contrast the various phase with LDA, PBEsol and PBE (solid, dashed, and dotted curves, respectively).
In the lower part of the panels, the intensity is also given on a log scale in order to ease the identification of weaker peaks.
The experimental results are shown as a black curve at the bottom of each panel.
The vertical dashed lines are drawn as a guide for the eyes at the frequencies measured experimentally (in black and brown to distinguish TO and LO frequencies).
}
\label{fig:raman}
\end{figure*}

Next, we compare the Raman spectra for the three CZTS phases calculated using LDA, PBEsol and PBE to experimental results,~\cite{dimitrievska2014multiwavelength} see Fig.~\ref{fig:raman}.
To take into account the polycrystalline nature of the experimental sample, we compute the Raman spectra for crystalline powders with a formalism that considers intensities in the parallel and perpendicular laser polarizations as prescribed by Caracas~\textit{et~al.}.~\cite{caracas2009elasticity} This provides an average over all the possible orientations of the crystal. In the present work we include the LO/TO splitting of the phonon frequencies~\footnote{All Raman tensors are computed including LO/TO contributions~\cite{veithen2005nonlinear} for light propagating along the c-axis of the conventional tetragonal cell.}, in contrast to \citet{skelton2015vibrational} where only the TO frequencies were considered. Raman intensities are computed in the backscattering geometry. The backscattering geometry and its convention are explained in section III of the Supplemental Material.~\cite{supplemental}
As discussed in Sec.~II, the PBEsol and PBE Raman spectra are obtained using their respective mode frequencies combined with the LDA intensities.
The calculated Raman spectra are simulated for an incident laser wavelength of 514~nm at 300~K (which corresponds to the experimental parameters~\cite{dimitrievska2014multiwavelength}) using a Lorentzian broadening with a FWHM of 5~cm$^{-1}$.
In Fig.~\ref{fig:raman}, we also present the Raman intensity on a logarithmic scale in order to ease the comparison for the modes with lower intensities.

Our PBEsol theoretical spectrum for KS is in reasonable agreement with previous calculations by Skelton~\textit{et~al.}~\cite{skelton2015vibrational} relying on finite differences.
There are however some discrepancies.
First, we observe a peak corresponding to the B(3) mode of KS that matches the experimental peak at 164~cm$^{-1}$ but which seems to be absent in Ref.~\cite{ skelton2015vibrational}
Furthermore, there are some differences in the intensity of some peaks.
In order to confirm our results, we have investigated this further by also using finite differences (see Supplemental Material~\cite{supplemental}).

As can be seen in Fig.~\ref{fig:raman}, the theoretical vibrational frequencies are red shifted by $\sim$10~cm$^{-1}$ when going from LDA to PBEsol and $\sim$15-20~cm$^{-1}$ more to PBE for all three phases, though it is more complicated than just a rigid shift.
A detailed quantitative analysis can be found in the Supplemental Material.~\cite{supplemental}
In contrast, the comparison with experiment is far from straightforward.
When considering the linear scale graphs, the general shape of the PMCA theoretical spectra seems to be most similar to the experimental one with essentially one main peak.
This is confirmed by our quantitative analysis (see Supplemental Material\cite{supplemental}).
However, when considering the logarithmic scale graphs (in which weaker peaks are emphasized), it is not the case anymore and our quantitative analysis shows that KS and ST match better the experimental results.
In any case, the agreement is far from perfect.
In fact, in the frequency region above 200~cm$^{-1}$, there are several peaks that are either observed in the experiment and missing in our calculated spectra of KS/ST/PMCA or vice versa.
While there are two LO modes [E(LO6) and B(LO6)] beyond 360~cm$^{-1}$ in the experimental spectrum, there are no LO modes observed in that range for any of the phases in our calculations.
This leads us to conclude that KS cannot be assumed to be the dominant phase while neglecting the presence of ST and PMCA, and most probably other phases.
This is in agreement with the conclusions by Khare~\textit{et~al.}.~\cite{ khare2012calculation}
Further work is thus needed considering other phases arising from possible cation disorder in the system.
Such a disorder, mainly in the Cu/Zn planes, has been shown to have considerable influence on the physical properties of the material also marked by a reduction in the band gap.~\cite{bourdais2016cu}
Note also that the general shape agreement found for PMCA points to a possible disorder also in the Cu/Sn planes.

Finally, we would like to point out that, for comparison with experiments, the Raman spectra have been presented using arbitrary units.
The theoretical results can however be compared on an absolute scale.
The absolute intensity of PMCA is found to be more than twice larger than that of KS or ST (see Supplemental Material~\cite{supplemental}).
Hence, its importance in the Raman spectra will be larger than its actual proportion in the samples.
This needs to be taken into account when comparing to the experimental spectra.

\subsection{Dielectric permittivity}

In this section, we present the calculated electronic ($\epsilon_{\infty}$) and static ($\epsilon_{0}$) dielectric permittivity tensors. 
These have two independent components $\epsilon_{\parallel}$ and $\epsilon_{\perp}$, respectively parallel and perpendicular to the $c$ axis.
The static dielectric tensor can be decomposed into the contribution of different modes (following the notations in Ref.~\onlinecite{gonze1997dynamical}):
\begin{equation}
\epsilon_{\alpha\beta}^{0}(\omega) = \epsilon_{\alpha\beta}^{\infty}+\sum_{m} \Delta\epsilon_{m,\alpha\beta}= \epsilon_{\alpha\beta}^{\infty} + \frac{4\pi}{\Omega_{0}}\sum_{m}\frac{S_{m,\alpha\beta}}{\omega_{m}^{2}}
\label{eqn:dielec-tensor}
\end{equation}
where $\Omega_{0}$ is the volume of the primitive unit cell.
$S_{m,\alpha\beta}$ is the mode-oscillator strength, is related to the eigendisplacements $U_{m}(\kappa\alpha)$ and Born effective charge tensor $Z^*$ by 
\begin{equation}
S_{m,\alpha\beta}=\Big( \sum_{\kappa\alpha^{'}} Z^{*}_{\kappa,\alpha\alpha^{'}}U_{m}^{*}(\kappa\alpha^{'}) \Big) \Big( \sum_{\kappa^{'}\beta^{'}} Z^{*}_{\kappa^{'},\beta\beta^{'}}U_{m}^{*}(\kappa^{'}\beta^{'}) \Big).
\end{equation}

The values of $\epsilon_{\parallel}$ and $\epsilon_{\perp}$ obtained for the three phases are reported in Table~\ref{tab:dielec-tensor}.
The dielectric tensors are all diagonal.
For all three structures, the results predicted by PBEsol lie in between the LDA and PBE ones. The value of $\epsilon_{\infty}$ in the $\perp$ direction is found to be greater than the $\parallel$ one for KS and PMCA while it is the opposite for PMCA. The dielectric tensor increases from KS to ST, and to PMCA. While the difference between the $\parallel$ and $\perp$ directions is typically $<$ 2 for KS and PMCA, there is a considerable difference ($>$ 4) for ST. 
To have more precise estimates of the dielectric tensor, a Hubbard $U$ correction on the Cu-$d$ states~\cite{persson2010electronic} or a hybrid functional calculation using HSE~\cite{paier2009cu} is required.
The modes that contribute most to the static dielectric constant in the direction parallel (resp. perpendicular) to the $c$ axis of the tetragonal cell are the $B(4)$ [resp. $E(4)$] modes for KS and the $B_2(3)$ [resp. $E(4)$] modes for ST and PMCA. A major contribution to the $B(4)$ and $B_2(3)$ modes originates from the S atoms in the $\parallel$ direction as seen in Fig.~\ref{fig:decomp}. In ST and PMCA, the remaining contribution is dominated by the $\parallel$ motion of Zn atoms while for KS it involves Cu atoms. Large eigendisplacements from these atoms increase their contribution to the dielectric tensor. For the $E(4)$ modes in KS/ST/PMCA, the major contribution is again from the S atoms which have a higher eigendisplacement, but this time in the $\perp$ direction. There is some contribution from the Cu and Zn atoms as well, while Sn and S in the $\parallel$ direction contribute to a lesser extent. The large eigendisplacements combined with the Born effective charges contribute to higher oscillator strength tensors ($S_{m,\alpha\beta}$) for each of the IR-active modes in KS, ST and PMCA. These values for the three different XC-functionals are given in Table S-I of the Supplemental Material.~\cite{supplemental}

\begin{table}[h]
\caption{
\label{tab:dielec-tensor}
Dielectric tensor computed with LDA, PBEsol and PBE. The contributions of the different phonon modes given as $\Delta\epsilon$ are calculated as in Equation~\ref{eqn:dielec-tensor}. The directions are indicated with respect to the $c$ axis of the conventional cell. The $\parallel$ and $\perp$ contributions are from the 6 $B$ and 6 $E$ modes in KS respectively; 4 $B_2$ and 6 $E$ modes in ST and PMCA respectively.
}
\begin{ruledtabular}
\begin{tabular}{llrrrrrrrrrrrrrr}
& & &\multicolumn{3}{c}{KS} & & & \multicolumn{3}{c}{ST} & & & \multicolumn{3}{c}{PMCA} \\
\cline{3-6}
\cline{8-11}
\cline{13-16}
& & &\multicolumn{1}{c}{$\parallel$} & & \multicolumn{1}{c}{$\perp$} & & & \multicolumn{1}{c}{$\parallel$} & & \multicolumn{1}{c}{$\perp$} & & & \multicolumn{1}{c}{$\parallel$} & & \multicolumn{1}{c}{$\perp$} \\
\hline
LDA  
&$\epsilon_{\infty}$  & & 10.67 & & 12.80 & & & 11.29 & & 13.35 & & & 15.63 & & 14.00 \\
&$\Delta\epsilon_{1}$ & &  0.34 & &  0.00 & & &  0.03 & &  0.09 & & &  0.05 & &  0.05 \\
&$\Delta\epsilon_{2}$ & &  0.04 & &  0.01 & & &  0.00 & &  0.02 & & &  0.01 & &  0.00 \\
&$\Delta\epsilon_{3}$ & &  0.02 & &  0.02 & & &  1.27 & &  0.00 & & &  0.87 & &  0.03 \\
&$\Delta\epsilon_{4}$ & &  1.60 & &  1.33 & & &  0.17 & &  1.82 & & &  0.74 & &  1.74 \\
&$\Delta\epsilon_{5}$ & &  0.28 & &  0.21 & & &       & &  0.50 & & &       & &  0.32 \\
&$\Delta\epsilon_{6}$ & &  0.40 & &  0.57 & & &       & &  0.46 & & &       & &  0.80 \\
\cline{2-16}
&$\epsilon_{0}$       & & 13.34 & & 14.95 & & & 12.76 & & 16.23 & & & 17.31 & & 16.94 \\
\hline 
PBEsol
&$\epsilon_{\infty}$   & & 10.80 & & 13.10 & & & 11.44 & & 14.41 & & & 16.05 & & 14.31 \\
&$\Delta\epsilon_{1}$  & &  0.30 & &  0.01 & & &  0.02 & &  0.05 & & &  0.07 & &  0.03 \\
&$\Delta\epsilon_{2}$  & &  0.04 & &  0.00 & & &  0.01 & &  0.04 & & &  0.05 & &  0.00 \\
&$\Delta\epsilon_{3}$  & &  0.03 & &  0.01 & & &  1.31 & &  0.00 & & &  0.83 & &  0.05 \\
&$\Delta\epsilon_{4}$  & &  1.75 & &  1.43 & & &  0.23 & &  1.98 & & &  0.86 & &  1.88 \\
&$\Delta\epsilon_{5}$  & &  0.30 & &  0.25 & & &       & &  0.58 & & &       & &  0.34 \\
&$\Delta\epsilon_{6}$  & &  0.48 & &  0.65 & & &       & &  0.56 & & &       & &  0.94 \\
\cline{2-16}
&$\epsilon_{0}$       & & 13.70 & & 15.46 & & & 13.01 & & 17.64 & & & 17.86 & & 17.56 \\
\hline
PBE
&$\epsilon_{\infty}$   & & 10.94 & & 13.49 & & & 11.63 & & 16.16 & & &  16.78 & & 15.00 \\
&$\Delta\epsilon_{1}$  & &  0.16 & &  0.04 & & &  0.00 & &  0.00 & & &  0.09 & &  0.00 \\
&$\Delta\epsilon_{2}$  & &  0.02 & &  0.03 & & &  0.13 & &  0.10 & & &  0.23 & &  0.01 \\
&$\Delta\epsilon_{3}$  & &  0.04 & &  0.01 & & &  1.41 & &  0.01 & & &  0.85 & &  0.07 \\
&$\Delta\epsilon_{4}$  & &  2.18 & &  1.57 & & &  0.29 & &  2.39 & & &  0.96 & &  2.23 \\
&$\Delta\epsilon_{5}$  & &  0.37 & &  0.45 & & &       & &  0.79 & & &       & &  0.48 \\
&$\Delta\epsilon_{6}$  & &  0.61 & &  0.75 & & &       & &  0.76 & & &       & &  1.17 \\
\cline{2-16}
&$\epsilon_{0}$        & & 14.32 & & 16.34 & & & 13.46 & & 20.21 & & & 18.92 & & 18.96 \\
\end{tabular}
\end{ruledtabular}
\end{table}

\section{Conclusions}
In this paper, we studied the structural, dynamical, dielectric, and static non-linear properties (non-resonant Raman scattering) of three different CZTS crystal structures (KS, ST, and PMCA) using DFPT.
In particular, we investigated the effect of various exchange correlation functionals (LDA, PBEsol, and PBE) on these properties.
In agreement with previous studies on many different materials, the structural properties obtained within PBEsol were found to be in between those calculated within LDA and PBE.
The dielectric properties and Born effective charge tensors showed a similar trend.
The computed LDA and PBEsol phonon frequencies presented a good agreement with those reported previously in the literature, with a red shift of $\sim$10~cm$^{-1}$ when moving from LDA to PBEsol.
The change from PBEsol to PBE lead to an extra redshift of $\sim$15-20~cm$^{-1}$ for all three phases.
None of the calculated transmittance and Raman spectra was found to coincide perfectly with the experimental one, though many similarities were evidenced.
This work hence points to the need to look beyond the standard phases (KS, ST, and PMCA) of CZTS by considering possible disorder effects on the vibrational properties. 
\begin{acknowledgments}

S.P.R. would like to thank the FRIA grant of Fonds de la Recherche Scientifique (F.R.S.-FNRS), Belgium; Y.G. and G.-M.R. are grateful to the F.R.S.-FNRS for the financial support. A.M. and G.-M.R. would like to acknowledge funding from the Walloon Region (DGO6) through the CZTS project of the ''Plan Marshall 2.vert'' program. Computational resources have been provided by the supercomputing facilities of the Universit\'{e} catholique de Louvain (CISM/UCL) and the Consortium des Equipements de Calcul Intensif en F\'{e}d\'{e}ration Wallonie Bruxelles (CECI) funded by the Fonds de la Recherche Scientifique de Belgique (FRS-FNRS).
\end{acknowledgments}

\bibliography{paper}
\bibliographystyle{apsrev4-1}

\end{document}